\documentclass{acm_proc_article-sp}
\usepackage{amssymb}
\usepackage{graphicx}
\usepackage{amsmath}
\usepackage{epsfig}
\usepackage{url}

\newtheorem{lemma}{Lemma}
\newtheorem{theorem}{Theorem}

\begin{document}

\title{On the Border Length Minimization Problem (BLMP) on a Square Array}

%\titlerunning{On the Border Length Minimization Problem}

%\author{Vamsi Kundeti \and Sanguthevar Rajasekaran \and Hieu Dinh}
%\institute{Department of Computer Science and Engineering\\
%University of Connecticut\\
%Storrs, CT 06269, USA\\
%\mailsa\\
%} 
\numberofauthors{3}
\author{
\alignauthor
Vamsi Kundeti\titlenote{Graduate student, Computer Science and Engineering Department}\\
	\affaddr{371 Fairfield Way, U-2031, A-38} \\
	\affaddr{Storrs, CT 06269}\\
	\email{vamsik@engr.uconn.edu}
\alignauthor
Sanguthevar Rajasekaran\titlenote{Dr. Rajasekaran, UTC Chair processor, Computer Science and Engineering Department} \\
	\affaddr{371 Fairfield Way, U-2155} \\
	\affaddr{Storrs, CT 06269}\\
	\email{rajasek@engr.uconn.edu}
\alignauthor
Hieu Dinh\titlenote{Graduate student, Computer Science and Engineering Department}\\
	\affaddr{371 Fairfield Way, U-2031, A-42} \\
	\affaddr{Storrs, CT 06269}\\
	\email{hdinh@engr.uconn.edu}
}
\maketitle
\begin{abstract}
Protein/Peptide microarrays are rapidly gaining momentum in the
diagnosis of cancer. High-density and high-throughput peptide arrays
are being extensively used to detect tumor biomarkers, examine
kinase activity, identify antibodies having low serum titers and
locate antibody signatures. Improving the yield of microarray fabrication 
involves solving a hard combinatorial optimization problem called the 
{\em Border Length Minimization Problem} $(BLMP)$. {\em An important
question that remained open for the past seven years is if the BLMP is tractable
or not}. We settle this open problem by proving that the BLMP is
$\cal NP$-hard. 
We also present a hierarchical refinement algorithm which can refine
any heuristic solution for the BLMP problem. We also prove that the 
TSP+1-threading heuristic is an $O(N)$-approximation.

The hierarchical refinement solver is available as an open-source
code at \url{http://launchpad.net/blm-solve}.
\end{abstract}

\category{F.2}{Analysis of Algorithms and Problem Complexity}[Complexity of proof procedures]
\category{G.2.2}{Graph Theory}[Graph Algorithms]
\category{F.1.3}{Complexity Measures and Classes}[Reducibility and completeness]
\terms{Border length minimization, Quadratic assignment, Microarray optimization, Appoximation algorithms,
Computational biology}
\section{Introduction} \label{sec:intro}

Cancer diagnosis research has taken a new direction recently by
adopting peptide microarrays for reliable detection of tumor
biomarkers (Chatterjee, et al.,~\cite{chatterjee2006}), (Melle, et
al.,~\cite{melle2004}), (Welsh, et al.,~\cite{welsh2003}). These
high-throughput arrays also find application in examining kinase
activity, identifying antibody signatures against tumor antigens,
etc. High-density peptide arrays are currently fabricated using
technologies such as photolithography or in-situ synthesis based on
micromirror arrays. The manufacturers of these arrays are facing
serious fabrication challenges due to unintended illumination effects
such as diffraction and scattering of light. These illumination
effects can be reduced dramatically by selecting a right placement
of the peptide probes before fabrication. Finding this placement can
be formulated as a combinatorial optimization problem, known as the
{\em Border Length Minimization Problem} (BLMP). Hannenhalli, et al.
first introduced BLMP in 2002 \cite{hannenhalli96}. Although the BLMP
was formulated in the context of DNA microarrays, peptide arrays
share a similar fabrication technology.

The BLMP can be stated as follows. Given $N^2$ strings of
the same length, how do we place them in a grid of size $N \times N$
such that the Hamming distance summed over all the pairs of neighbors
in the grid is minimized? The BLMP has received a lot of attention from
many researchers. The earliest algorithm suggested by Hannenhalli,
et al. reduces BLMP to TSP (Traveling Salesman Problem) by computing
a tour of the strings and then threading the tour on the grid
\cite{hannenhalli96}. Kahng, et al. have proposed
several other heuristic algorithms which are considered the best
performing algorithms in practice \cite{epitaxial03}. De Carvalho,
et al. introduced a quadratic program formulation of the BLMP but
unfortunately the quadratic program is an intractable problem
\cite{CAR-RAH-2006a}.
Later, Kundeti and Rajasekaran formulated the problem as an integer
linear program which performs better than the quadratic program in
practice \cite{kundetibibe2009}.

Despite many studies on the BLMP, the question of whether BLMP is
tractable or not remained open for the past 7 years. In this paper,
we show that the BLMP is $\cal NP$-hard. We also consider a
generalization of the BLMP called the \emph{Hamming Graph Placement
Minimization Problem} (HGPMP). We show that some special cases of
the HGPMP are also $\cal NP$-hard. On the algorithmic side, we show
that a simple version of the algorithm suggested by Hannenhalli, et
al. is an $O(N)$-approximation. On the practical side, we propose a
refinement algorithm which takes any solution and tries to improve
it. An experimental study of this refinement algorithm is also
included.

Our paper is organized as follows. Section \ref{sec:definitions}
formally defines the BLMP and HGPMP. Section \ref{sec:NP-hard}
provides the $\cal NP$-hardness proof of the BLMP and some special cases of
the HGPMP. Section \ref{sec:approx} gives the $O(N)$-approximation
algorithm and the refinement algorithm for the BLMP. Section
\ref{sec:experiment} provides an experimental evaluation of the
refinement algorithm. Finally, Section \ref{sec:conclusion}
concludes our paper and discusses some open problems.

\section{Problem definition} \label{sec:definitions}

Let $S$ be a set of strings of the same length with $S=\{s_1,\dots,
s_{n}\}$ and let $G=(V,E)$ be a graph with $|V| = n$. A
placement of $S$ on $G$ is a bijective map $f: S \rightarrow V$. Let
$f^{-1}(u)$ be the string that is mapped to vertex $u$ by the placement
$f$. We denote the Hamming distance between two strings $s_i$ and
$s_j$ as $\delta(s_i,s_j)$. The cost of placement $f$ is $Cost(f)
= \sum_{e=(u,v)\in E} \delta(f^{-1}(u),f^{-1}(v))$. The Hamming Graph
Placement Minimization Problem (HGPMP) is defined as follows. Given
$S$ and $G$, find a placement of $S$ on $G$ of minimum cost.
We denote the optimal
cost as $OPT(S,G)$, or simply as $OPT$ if it is clear what $S$ and $G$ are.

Obviously, if $G$ is a ring graph, then HGPMP is the same as the
well-known Hamming Traveling Salesman Problem (HTSP). If $G$ is a
grid graph of size $N \times N$ (where $N^2 = n$), then HGPMP
becomes the Border Length Minimization Problem (BLMP), which is the
main study of our paper.

\section{$\cal NP$-hardness of the BLMP \\ and HGPMP} 
\label{sec:NP-hard}

\begin{theorem}\label{maintheorem}
The BLMP is $\cal NP$-hard.
\end{theorem}

We will show that the Hamming traveling salesperson problem (HTSP)
for strings (with the Hamming distance metric) polynomially reduces
to the BLMP. The HTSP is already defined in Section
\ref{sec:definitions}.

The idea of the proof is that given $4N$ strings for the HTSP we
construct $(N + 1)^2$ strings for the BLMP such that from an optimal
solution to this BLMP, we can easily obtain an optimal solution for
the HTSP. So we need to consider the variant of the HTSP in which
the number of strings is divisible by $4$. The proof will be
presented in stages. The next three subsections present some
preliminaries needed for the proof of the theorem. Followed by these
subsections, the proof is presented.

\subsection{$4N$-strings traveling salesperson problem}
Define an instance of the HTSP as a {\em $4N$-strings
HTSP} if the number of strings in the input is $4N$ (for some integer $N$).
In this section we show that the $4N$-strings
HTSP is $\cal NP$-hard.

\begin{theorem}\label{4N-HTSP}
$4N$-strings HTSP is $\cal NP$-hard.
\end{theorem}

\noindent{\bf Proof:} We will show that the HTSP polynomially
reduces to the $4N$-strings HTSP. Let $S=\{s_1,s_2,\ldots,s_n\}$ be
the input for any instance of the HTSP. Let $\ell$ be the length of
each input string. Append a string of $2n\ell$ $\overline{0}$'s to
the left of each $s_i$ to get $s_i^\prime$ (for $1\leq i\leq
(n-1))$. For example, if $n=4,\ell=3$ and $s_1=\overline{101}$, then
$s_1^\prime$ will be $\overline{000000000000000000000000101}$. We
append $2n\ell$ $\overline{1}$'s to the left of $s_n$ to get
$s_n^\prime$. We will generate an instance $S'$ of the $4N$-strings
HTSP that has as input $4N$ strings, where $N = \lceil \frac{n}{4}
\rceil$. $S'$ will have
$s_1^\prime,s_2^\prime,\ldots,s_{n-1}^\prime$ and $1, 2, 3$ or $4$
copies of $s_n^\prime$ depending on whether $n=4N, 4N-1, 4N-2$, or
$4N-3$, respectively.

It is easy to see that in an optimal tour for the above
$4N$-strings HTSP instance, all the copies of $s_n^\prime$ will be successive
and that an optimal solution for $S$ can be obtained readily from an
optimal solution for $S'$. $\Box$

\subsection{A special instance of the BLMP}
Consider the following $(N+1)^2$ strings as an input for
the BLMP:\protect\linebreak
$t_1, t_2, \dots, t_{4N}, t, t, \dots, t$. Here there are $N^2-2N+1$
copies of $t$. There is a positive integer $k$ such that
$\delta(t_i,t)=k$ for any $1 \leq i\leq 4N$ and $2k \geq
\delta(t_i,t_j)
> \frac{7}{4}k$ for any $1 \leq i \neq j \leq 4N$.

\begin{figure}
\begin{center}
  \includegraphics[width=2in]{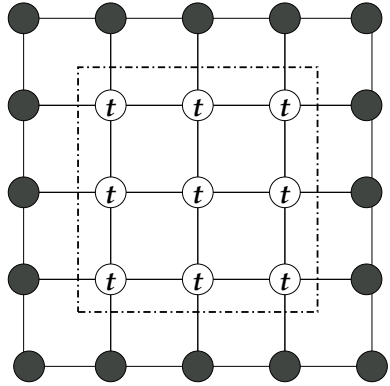}\\
  \caption{An illustration for Lemma \ref{specialcase} with $N=4$. Each $t_i$ lies on a dark vertex in the grid.}\label{fig:blmp}
\end{center}
\end{figure}

\begin{lemma}\label{specialcase}
In any optimal solution to the above BLMP instance,
$t_1,t_2,\ldots,t_{4N}$ will lie on the boundary of the $(N+1)\times
(N+1)$ grid (see Figure \ref{fig:blmp}).
\end{lemma}

\noindent{\bf Proof:} This can be proven by contradiction. Let $T$
be the collection of the strings $t_1,t_2,\ldots,t_{4N}$. Let $q$ be
one of the strings from $T$ that has a degree of 4 in an optimal
placement. Let $r$ be one of the strings equal to $t$ that lies in
the boundary. Next we show that we can get a better solution by
exchanging $q$ and $r$.

Let $u$ be the number of neighbors of $q$ from $T$. Let $v$ be the
number of neighbors of $r$ from $T$. Note that $0\leq u\leq 4$ and
$0\leq v\leq 3$. In the current solution, the total cost incurred by
$q$ and $r$ is at least $\frac{7}{4}ku + k(4-u) + kv=\frac{3}{4}ku +
kv + 4k$. If we exchange $q$ and $r$, the new total cost incurred by
$q$ and $r$ is strictly less than $ku + 2kv + k(3-v)=ku + kv + 3k$.
The old cost minus the new cost is strictly greater than $k -
\frac{1}{4}ku \geq 0$.

We thus conclude that all the strings of $T$ lie on the boundary of
the grid in any optimal solution. $\Box$

\subsection{A special set of strings and some operations on strings}\label{secstrings}
We denote the (ordered) concatenation of two strings $x$ and $y$ as
$x + y$. If $x$ and $x'$ (respectively $y$ and $y'$) have the same
length then, clearly, $\delta(x+y,x'+y') = \delta(x,x') + \delta(y,y')$.

Given a string $x=\overline{x_1x_2\dots x_l}$ and an integer $h$,
let $REP_h(x)$ be the string $\overline{x_1x_1\dots x_1 x_2x_2\dots
x_2\dots x_lx_l\dots x_l}$, where each $x_i$ appears $h$ times
($REP$ stands for ``replicate''). It is not hard to see that if $x$
and $y$ have the same length, then $\delta(REP_h(x),REP_h(y)) =
h\delta(x,y)$.

Given an integer $n$, we can construct a set of $n$ strings of
length $n$ each, $A_n=\{a_1,a_2,\dots,a_n\}$, such that
$\delta(a_i,a_j)=2$ for any $1 \leq i \neq j \leq n$. One way to
construct $A_n$ is to let $a_i=\overline{00\dots0100\dots0}$, where
there are $(i-1)$ $\overline{0}$'s before $\overline{1}$. It is easy
to check that $\delta(a_i,a_j)=2$ for any $1 \leq i \neq j \leq n$.

\subsection{Proof of the main theorem}
Now we are ready to present the proof of Theorem \ref{maintheorem}.
Let $S=\{s_1,s_2,\ldots,s_{4N}\}$ be the input for any instance of
the $4N$-strings HTSP. Each $s_i$ has the length $l$. We will
generate $(N+1)^2$ strings such that an optimal solution for the
BLMP on these $(N+1)^2$ strings will yield an optimal solution for
the $4N$-strings HTSP on $S$.

The input for the BLMP instance that we generate will
be\protect\linebreak $T=\{t_1,t_2,\ldots,t_{4N},t,t,\ldots,t\}$
where $t$ occurs $N^2-2N+1$ times. We set $t_i = REP_h(a_i) +
REP_2(s_i)$, where $a_i$ is the $i$-th string in the set $A_{4N}$
defined in subsection \ref{secstrings}. We will choose $h$ later.
Also, we set $t = REP_{4Nh}(\overline{0}) + \overline{0101\dots01}$,
where the string $\overline{01}$ is repeated $l$ times. We can
easily check that:
\begin{eqnarray}
\delta(t_i,t) & = & h + l \,\,\mbox{ for any } 1 \leq i \leq 4N \label{eq01} \\
\delta(t_i,t_j) & = & 2h + 2\delta(s_i,s_j) \leq 2h + 2l \label{eq02}\\
&&\mbox{ for any } 1 \leq i \neq j \leq 4N \nonumber
\end{eqnarray}

We choose $h$ so that $T$ satisfies the condition in Lemma
\ref{specialcase}. Particularly, choose $h = 8l$. Now we will show
that $OPT_{BLMP}(T) = 4(N-1)(h+l) + 8Nh + 2OPT_{HTSP}(S)$, which in turn
means that an optimal solution for the BLMP on $T$ will yield an optimal
solution for the $4N$-strings HTSP on $S$.

Let $A = s_{i_1},s_{i_2},\dots,s_{i_{4N}}$ be an optimal tour for
the $4N$-string HTSP on $S$. We construct a solution $A'$ for the
BLMP on $T$ by placing $t_i$'s on the border of the grid in the
order $t_{i_1},t_{i_2},\dots,t_{i_{4N}}$ and placing the copies of
$t$ on the center of the grid. By the equalities (\ref{eq01}) and
(\ref{eq02}), the cost of $A'$ is $Cost(A')=4(N-1)(h+l) + 8Nh +
2Cost(A)$. Therefore, $OPT_{BLMP}(T) \leq 4(N-1)(h+l) + 8Nh +
2OPT_{HTSP}(S)$.

On the other hand, let $B$ be an optimal solution for the BLMP on
$T$. By Lemma \ref{specialcase}, $t_i$'s lie on the border of the
grid and the copies of $t$ lie on the center of the grid. Assume that
$t_i$'s lie in the order $t_{i_1},t_{i_2},\dots,t_{i_{4N}}$. We can
construct a tour $B'$ for the $4N$-strings HTSP on $S$ in the order
$s_{i_1},s_{i_2},\dots,s_{i_{4N}}$. By the equalities (\ref{eq01})
and (\ref{eq02}), $Cost(B)=4(N-1)(h+l) + 8Nh + 2Cost(B')$. Hence,
$OPT_{BLMP}(T) \geq 4(N-1)(h+l) + 8Nh + 2OPT_{HTSP}(S)$.

This completes the proof of Theorem \ref{maintheorem}. $\Box$

\subsection{$\cal NP$-hardness of the HGPMP for other special cases}

We can generalize the result in Theorem \ref{maintheorem} for other
special cases of the HGPMP. We say graph $G$ is
\emph{``bordered-ring"} if $G$ is undirected and $G$ has a ring of
size $\Omega(n^\alpha)$ for some constant $\alpha
> 0$ such that every vertex in the ring has degree no greater than
$d$ and every vertex outside the ring has degree greater than $d$
for some $d \geq 3$. For grid graphs, $\alpha=\frac{1}{2}$ and
$d=3$. Some variants of grid graphs like Manhattan grids are
bordered-ring as well.

\begin{theorem}
The HGPMP is $\cal NP$-hard even if $G$ is bordered-ring.
\end{theorem}

\noindent{\bf Proof:} By a similar
reduction to that of the BLMP above, the theorem follows. $\Box$

\subsection {An alternate $\cal NP$-hardness proof \\ for the BLMP}

In this section, we give an alternate $\cal NP$-hardness proof for the BLMP by
showing that another variant of the HTSP called $k$-Segments HTSP
polynomially reduces to the BLMP. We believe that the techniques introduced
in both of our proofs will find independent applications.

\subsubsection{$k$-Segments traveling salesperson problem} ~\\
We define the $k$-segments HTSP and show that it is NP-hard.
Consider an input of $n$ strings: $s_1,s_2,\ldots,s_n$. The problem
of $k$-segments HTSP is to partition the $n$ strings into $k$ parts
such that the sum of the optimal tour costs for the individual parts
is minimum.

\begin{theorem}\label{ksTSP}
The $k$-segments HTSP for strings is $\cal NP$-hard.
\end{theorem}

\noindent{\bf Proof:} We will prove this for $k=4$ (since this is
the instance that will be useful for us to prove the main result)
and the theorem will then be obvious.

We will show that the HTSP polynomially reduces to the $4$-segments
HTSP. Let $S=\{s_1,s_2,\ldots,s_n\}$ be the input to any instance of
the HTSP. We will generate an instance of the $4$-segments HTSP that
has as input $(n+3)$ strings. Let $l$ be the length of each string
in $S$. Note that the optimal cost for the HTSP with input $S$ is
$\leq nl$.

Consider the 4 strings:
$\overline{1110},~\overline{1101},~\overline{1011},~\overline{0111}$.
The distance between any two of them is $2$. Now replace each
$\overline{1}$ in each of these $4$ strings with a string of $nl$
$\overline{1}$'s. Also, replace each $\overline{0}$ in each of these
strings with a string of $nl$ $\overline{0}$'s. Call these new
strings $t_1,t_2,t_3,t_4$. The distance between any two of these
strings is $2nl$.

The input strings for the $4$-segments HTSP are
$q_1,q_2,\ldots,q_{n+3}$ and are constructed as follows: $q_i$ is
nothing but $s_i$ with $t_1$ appended to the left, for $1\leq i\leq
n$. $q_{n+1}$ is a string of length $4nl+l$ whose $l$ LSBs are
$\overline{0}$'s and whose $4nl$ MSBs equal $t_2$. $q_{n+2}$ is a
string of length $4nl+l$ whose $l$ LSBs are $\overline{0}$ and whose
$4nl$ MSBs equal $t_3$. Also, $q_{n+3}$ has all $\overline{0}$'s in
its $l$ LSBs and its $4nl$ MSBs equal $t_4$.

Clearly, in an optimal solution for the $4$-segments HTSP instance,
the four parts have to be
$\{q_1,q_2,\ldots,q_n\},~\{q_{n+1}\},~\{q_{n+2}\}$, and
$\{q_{n+3}\}$. As a result, we can get an optimal solution for the
HTSP instance given an optimal solution for the $4$-segments HTSP
instance. $\Box$

\subsubsection{A special instance of the BLMP} ~\\
Consider the following $n^2$ strings as an input for the
BLMP:\protect\linebreak $t_1,t_2,\ldots,t_{n},t,t,\ldots,t$. Here
there are $n^2-n$ copies of $t$. Also, $\delta(t_i,t_j)=16$ for any
$i$ and $j$ less than or equal to $n$. $\delta(t_i,t)=9$ for any
$i\leq n$.

\begin{lemma}\label{specialcase-alter}
In an optimal solution to the above BLMP instance,
$t_1,t_2,\ldots,t_n$ lie on the boundary of the $n\times n$ grid and
moreover these strings are found in four segments of successive
nodes.
\end{lemma}

\noindent{\bf Proof:} Let $T$ be the collection of strings
$t_1,t_2,\dots,t_n$. By Lemma \ref{specialcase}, we conclude that
all the strings of $T$ lie on the boundary of the grid in an optimal
solution.

Let $S_1$ and $S_2$ be two segments such that $S_1$ and $S_2$
consist of strings from $T$, strings in $S_1$ are in successive
nodes, strings in $S_2$ are in successive nodes, and these two
segments are not successive. Consider the case when none of these
strings is in a corner of the grid. Let
$S_1=\{a_1,a_2,\ldots,a_{n_1}\}$ and
$S_2=\{b_1,b_2,\ldots,b_{n_2}\}$. Let
$C(S_1)=\sum_{i=1}^{n_1-1}\delta(a_i,a_{i+1})$ and
$C(S_2)=\sum_{i=1}^{n_2-1}\delta(b_i,b_{i+1})$. The total cost for
these two segments is $C(S_1)+C(S_2)+9(n_1+n_2)+36$. If we join
these two segments into one, the new cost will be
$C(S_1)+C(S_2)+9(n_1+n_2)+34$.

Thus it follows that all the strings of $T$ will be on the boundary
and they will be found in successive nodes in any optimal solution.
Also it helps to utilize the corners of the grid since each use of a
corner will reduce the total cost by 9. Therefore in an optimal
solution there will be four segments such that all the segments are
in the boundary of the grid, each segment has strings from $T$ in
successive nodes, and one string of each segment occupies a corner
of the grid. In other words, an optimal solution for the BLMP
instance contains an optimal solution for the 4-segments TSP
corresponding to $T$. The optimal cost for this BLMP instance is
$25n-28$. $\Box$

\subsubsection{Construction of strings for the above BLMP
instance}\label{secstrings-alter} ~\\ We can construct $n^2$ strings
that have the same properties as the ones in the above BLMP
instance.

To begin with, we construct $(n+1)$ binary strings of length $n$
each. The string $t_i$ has all $\overline{1}$'s except in position
$i$, for $1\leq i\leq n$. The position of the LSB of any string is
assumed to be $\overline{1}$. String $t_{n+1}$ has all
$\overline{1}$'s. Clearly, $\delta(t_i,t_j)=2$ for any $i$ and $j$
less than or equal to $n$. Also, $\delta(t_i,t_{n+1})=1$ for any
$1\leq i\leq n$.

Now, in each $t_i$ (for $1\leq i\leq (n+1)$) replace every
$\overline{1}$ with a string of eight $\overline{1}$'s and replace
each $\overline{0}$ with a string of eight $\overline{1}$'s. After
this change, $\delta(t_i,t_j)=16$ for any $1\leq i,j\leq n$ and
$\delta(t_i,t_{n+1})=8$ for any $1\leq i\leq n$.

Finally, append a $\overline{0}$ to the left of each $t_i$ (for
$1\leq i\leq n$) as the MSB. Also, append a $\overline{1}$ to the
left of $t_{n+1}$. In this case, $\delta(t_i,t_j)=16$ for any $1\leq
i,j\leq n$ and $\delta(t_i,t_{n+1})=9$ for any $1\leq i\leq n$.

\subsubsection{The alternate proof of the main theorem}~\\
Let $S=\{s_1,s_2,\ldots,s_n\}$ be the input for any instance of the
HTSP. We will generate $n^2$ strings such that an optimal solution
for the BLMP on these $n^2$ strings will yield an optimal solution
for the $4$-segments HTSP on $S$.

We will use as the basis the $(n+1)$ strings generated in the above
section. Recall that these strings $t_1,t_2,\ldots,t_{n+1}$ are of
length $(8n+1)$ each. Also, $\delta(t_i,t_j)=16$ for any $1\leq
i,j\leq n$ and $\delta(t_i,t_{n+1})=9$ for any $1\leq i\leq n$.

Replace each $\overline{0}$ in each of the above strings with $nl$
$\overline{0}$'s and replace each $\overline{1}$ in each of these
strings with $nl$ $\overline{1}$'s. Now, $\delta(t_i,t_j)=16nl$ for
any $1\leq i,j\leq n$ and $\delta(t_i,t_{n+1})=9nl$ for any $1\leq
i\leq n$. Each of these strings is of length $(8n+1)nl$.

Replace each $\overline{0}$ in each $s_i$ with two $\overline{0}$'s
(for $1\leq i\leq n$) and replace each $\overline{1}$ in each $s_i$
with two $\overline{1}$'s and let $s^\prime_i$ be the resultant
string.  Note that an optimal solution for the $4$-segments HTSP on
the revised $S$ will also be an optimal solution for the
$4$-segments HTSP on the old $S$. If $l$ is the length of each
string in the old $S$, then $2l$ will be the length of each revised
input string.

The input for the BLMP instance that we generate will be
$q_1,q_2,\ldots,q_n,t,t,\ldots,t$ where $t$ occurs $n^2-n$ times.
Each of these strings will be of length $(8n+1)nl+2l$. The string
$q_i$ will have $s^\prime_i$ in its $2l$ LSBs and it will have $t_i$
in its $(8n+1)nl$ MSBs, for $1\leq i\leq n$. The string $t$ will
have $t_{n+1}$ in its $(8n+1)nl$ MSBs. Its $2l$ LSBs will be
$\overline{0101\ldots 01}$, i.e., the string $\overline{01}$ is
repeated $l$ times. Note that
$\delta(q_i,q_j)=16nl+\delta(s^\prime_i,s^\prime_j)$ for any $1\leq
i,j\leq n$. Also, $\delta(q_i,t)=9nl+l$ for any $1\leq i\leq n$.

Note that strings of this BLMP instance are comparable to the
strings we had for Lemma \ref{specialcase-alter}. This is because
the interstring distances are very nearly in the same ratios for the
two cases. As a result, using a proof similar to that of Lemma
\ref{specialcase-alter}, we can show that the strings
$t_1,t_2,\ldots,t_n$ will all lie in the boundary of the grid in an
optimal solution to the above BLMP. Let $T=\{t_1,t_2,\ldots,t_n\}$.
Also, the strings of $T$ will be found in four segments such that
one string of each segment occupies one of the corner nodes of the
grid. Let $S_1,S_2,S_3,$ and $S_4$ stand for the strings in these
four segments, respectively. Let $C_1,C_2,C_3$, and $C_4$ be the
optimal tour costs for $S_1,S_2,S_3$, and $S_4$, respectively.

Let $|S_i|=n_i$ for $1\leq i\leq 4$. The total cost (i.e., the
border length) for the above BLMP solution can be computed as
follows. Consider $S_1$ alone. The cost due to this segment is
$C_1+2(9nl+l)+(n_1-1)(9nl+l)$. The cost $2(9nl+l)$ is due to the two
end points of the segment $S_1$. The cost $(n_1-1)(9nl+l)$ is due to
the fact that each string of $S_1$ (except for the one in a corner
of the grid) is a neighbor of a $t$. Upon simplification, the cost
for $S_1$ is $C_1+(n_1+1)(9nl+l)$. Summing over all the four
segments, the total cost for the BLMP solution is
$C_1+C_2+C_3+C_4+(n+4)(9nl+l)$. The minimum value of this is
obtained when $S_1,S_2,S_3$, and $S_4$ form a solution to the
$4$-segments HTSP on $T$.

Clearly, an optimal solution for the $4$-segments HTSP on $T$ will
also yield an optimal solution for the $4$-segments HTSP on $S$.
This can be seen as follows. Consider the strings in $S_i$ and let
$Q_i=a^i_1,a^i_2,\ldots,a^i_{n_i}$ be the corresponding input
strings (of $S$), for $1\leq i\leq 4$. Note that $C_i$ is nothing
but $(n_i-1)(16nl)$ plus twice the optimal tour cost for $Q_i$, for
$1\leq i\leq 4$. Thus, $C_1+C_2+C_3+C_4$ is equal to
$(n-4)16nl+2(C^\prime_1+C^\prime_2+C^\prime_3+C^\prime_4)$ where
$C^\prime_i$ is the optimal tour cost for $Q_i$, for $1\leq i\leq
4$.

This completes the proof of Theorem \ref{maintheorem}. $\Box$

\section{Algorithms for the BLMP} \label{sec:approx}

\subsection{An $O(N)$-approximation algorithm}

\begin{figure}
\begin{center}
  \includegraphics[width=2in]{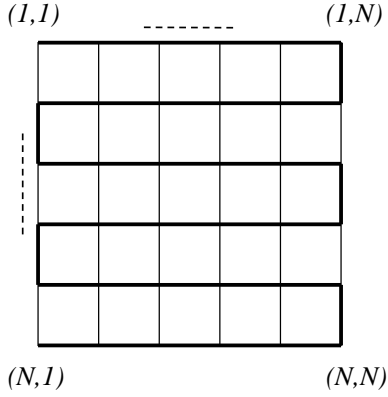}\\
  \caption{The thick dark line corresponds to an optimal tour on the input strings}\label{fig:approx}
\end{center}
\end{figure}

In this section, we will show that a simple version of the algorithm
suggested by Hannenhalli, et al. is actually an $O(N)$-approximation
algorithm. This algorithm can be described as follows. Assume that
the input is the set of strings $S =\{s_1,s_2,\dots,s_{N^2}\}$. The
algorithm first computes a tour $T$ on strings in $S$. Then it
threads the tour $T$ into the grid in row-major order (see Figure
\ref{fig:approx}). The first step can be done by calling the
$\frac{3}{2}$-approximation algorithm for the HTSP suggested by
\cite{1.5-approx-HTSP}.

\begin{lemma}\label{lemma:approx}
$OPT_{HTSP}(S) \leq 2OPT_{BLMP}(S)$.
\end{lemma}

\noindent{\bf Proof:} Let $A$ be an optimal solution for the BLMP on
$S$. Consider the path $P'$ drawn as the thick dark line in Figure
\ref{fig:approx}. Obviously, $Cost(P') \leq Cost(A)=OPT_{BLMP}(S)$.
Let $s_{i_1}$ and $s_{i_{N^2}}$ be the two endpoints of $P'$. Since
the Hamming distance satisfies the triangular inequality,
$\delta(s_{i_1},s_{i_{N^2}}) \leq Cost(P')$. Consider the tour that
starts at $s_{i_1}$, traverses along the path $P'$ to $s_{i_{N^2}}$
and comes back to $s_{i_1}$. Obviously, the cost of the tour is
$Cost(P')+\delta(s_{i_1},s_{i_{N^2}}) \leq 2Cost(P') \leq 2Cost(A)$.
Hence, $OPT_{HTSP}(S) \leq 2OPT_{BLMP}(S)$. $\Box$

\begin{theorem}\label{theorem:approx}
The above algorithm yields an $O(N)$-approximate solution.
\end{theorem}

\noindent{\bf Proof:} First, we see that $Cost(T) \leq
\frac{3}{2}OPT_{HTSP}(S) \leq 3OPT_{BLMP}(S)$. The first inequality
is due to the $\frac{3}{2}$-approximation for the HTSP. The second
inequality is due to Lemma \ref{lemma:approx}. Now let us analyze
the cost of the solution $F$ produced by the algorithm. Consider the
path $P$ drawn as the thick dark line in Figure \ref{fig:approx}.
Obviously, $Cost(P) \leq Cost(T)$. Also, the total cost of the $N$
rows in $F$ is no more than $Cost(P)$. By the triangle inequality,
it is not hard to see that the cost of each column in $F$ is no more
than $Cost(P)$. Therefore, $Cost(F) \leq (N+1)Cost(P) \leq
(N+1)Cost(T)\leq 3(N+1)OPT_{BLMP}(S) = O(N)OPT_{BLMP}(S)$. $\Box$

\subsection{A hierarchical refinement algorithm}
\label{sec:hier}
\begin{figure}
\begin{center}
\includegraphics[scale=0.5]{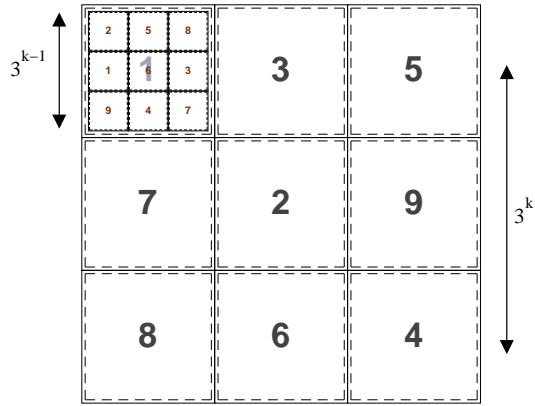}
\end{center}
\caption{Illustration of the hierarchical refinement algorithm with degree of refinement $3$. This shows
the possible optimal solutions (i.e. permutation among sub-problems) at the top-most and penultimate levels}
\label{fig:hier_blm}
\end{figure}

Several heuristics such as the Epitaxial growth have been proposed to solve the BLMP problem earlier. 
However most of these heuristics do not improve the cost monotonically. Local search based 
algorithms are often employed to solve hard combinatorial problems. We now introduce a 
hierarchical refinement algorithm ($HRA$). This refinement technique can be applied 
to any heuristic placement to refine the cost and get a better placement. Let $N$ be the number
of probes in the placement, $d$ a positive integer such that $d^x=N, x\geq 1$ is called the
degree of refinement. The refinement algorithm starts with a given placement, then it divides 
the placement into $s^0_1,s^0_2\ldots s^0_{N/d^2}$ sub-problems with $d^2$ probes per sub-problem. 
Each of these sub-problems is solved optimally -- an optimal permutation among the probes
is found. After this every $d^2$ sub-problems are combined into a new sub-problem 
$s^1_i = \displaystyle\cup_{j=1}^{d^2} s^0_{id^2+j}, 1\leq i \leq N/d^3$. To solve 
$s^1_i$ optimally we identify an optimal permutation among $s^0_{id^2 +j} \in s^1_i,1\leq j \leq d^2$.
This process continues until we are left with no sub-problems to solve. See Figure~\ref{fig:hier_blm}.

We should remark that while solving a sub-problem optimally, we also consider the cost contributed from the 
neighboring sub-problems. This ensures the monotonic improvement in the placement cost. The refinement algorithm 
asymptotically runs in $\Theta(d^2!N)$ time. If $d=O(1)$, the refinement algorithm 
runs in linear time. For small values of $d$, the algorithm performs well in practice. $HRA$ is a
deterministic refinement algorithm. We
further extend this by introducing randomness. The Randomized Hierarchical Refinement Algorithm ($RHRA$)
is similar to the $HRA$ algorithm. $RHRA$ randomly selects a sub-square within the given placement
and applies the $HRA$ technique to the selected sub-square. Similar to local search algorithms, repeating
$RHRA$ algorithm several times improves the placement cost monotonically. We study the performance of both these
algorithms in section~\ref{sec:experiment}.

\subsection{Quad epitaxial algorithm}
\label{sec:quad_epi}
The {\em epitaxial} ($EPX$) placement suggested in~\cite{epitaxial03} places a randomly selected probe at the
center of the array, it continues placing the probes greedily around the locations adjacent to the placed
probes to minimize the cost (i.e. the algorithm almost spends $O(N^2)$ time to place each probe). The 
epitaxial algorithm gives good results for small arrays but for larger arrays the epitaxial algorithm is 
impractical and extremely slow. We propose the Quad Epitaxial ($QEPX$) algorithm as a simple extension to the epitaxial
algorithm. $QEPX$ yields good performance and is very fast compared to the $EPX$ algorithm. The basic
idea behind the $QEPX$ algorithm is to divide the array into four parts, apply $EPX$ algorithm for each
of the four parts and finally find an optimal arrangement among the four parts. In section~\ref{sec:experiment}
we compare the $QEPX$ algorithm with $EPX$ algorithm.

\section{Experimental study} 
\label{sec:experiment}
\subsection{Performance of the $QEPX$ algorithm}
In this section we compare the performance of $QEPX$ algorithm introduced earlier. We use randomly generated 
probe arrays of size $32^2$,$64^2$,$128^2$ and $256^2$. In all of our experimental studies we compute a
{\em lower bound} on the solution by picking the smallest $2N(N-1)$ edges from the complete Hamming distance graph.
Column-$4$(INIT COST) in the table~\ref{table:quad_epx} indicates the placement cost obtained by placing the probes 
in the row major order as given by the input. Column-$5$($8$) indicates the final placement cost obtained by the 
epitaxial (quad) algorithm. As we can see from columns $7$ and $10$, the refinement obtained by the $QEPX$ algorithm 
is very close to the $EPX$ algorithm. On the other hand $QEPX$ runs $3.6X$ faster than the $EPX$ algorithm. As we can 
see from table~\ref{table:quad_epx}, as the chip size increases $EPX$ algorithm becomes very slow. We ran both 
$EPX$ and $QEPX$ algorithms on a chip size of $243\times 243$ with a time limit of $60$ minutes. The $QEPX$ algorithm 
took around $12$ minutes to complete and improved the input placement cost by $36$\%. On the other hand the $EPX$ 
algorithm did not complete the placement. From our experiments we conclude that the $QEPX$ can provide a good placement 
which we can use as an input for refinement/local search algorithms such as $RHRA$. In the next sub-section we provide 
our experimental study of $HRA$ and $RHRA$ algorithms on various placement heuristics.
\begin{table*}
\begin{center}
\begin{tabular}{|l|l|l|l|l|l|l|l|l|l|}
\hline 
&&&&&&&&&\\
TEST&PROBES & LOWER & INIT & EPX & TIME & REFINED &QEPX & TIME & REFINED\\
CASE&&BOUND&COST&&(sec)&PRECENT&&(sec)&PRECENT\\
&&&&&&&&&\\
\hline 
t-0 & 1024 & 23480 & 37192 & 27591 & 0.60 & 25.81\% & 28060 & 0.42 & 24.55\% \\ 
\hline 
t-1 & 1024 & 23427 & 37029 & 27472 & 0.62 & 25.81\% & 28151 & 0.43 & 23.98\% \\ 
\hline 
t-0 & 4096 & 86818 & 151116 & 106471 & 10.70 & 29.54\% & 107805 & 3.05 & 28.66\% \\ 
\hline 
t-1 & 4096 & 86897 & 151176 & 106430 & 10.37 & 29.60\% & 107634 & 3.23 & 28.80\% \\ 
\hline 
t-0 & 16384 & 322129 & 609085 & 410301 & 180.00 & 32.64\% & 411746 & 43.93 & 32.40\% \\ 
\hline 
t-1 & 16384 & - & 608928 & 409625 & 185.88 & 32.73\% & 410902 & 44.70 & 32.52\% \\ 
\hline 
t-0 & 65536 & - & 2447885 & 2447885 & - & 0.00\% & 1563369 & 765.79 & 36.13\% \\ 
\hline 
t-1 & 65536 & - & 2427143 & 2427143 & - & 0.00\% & 1562630 & 774.33 & 35.62\% \\ 
\hline 
\end{tabular}

\end{center}
\caption{Comparison between {\em epitaxial} and {\em quad epitaxial} }
\label{table:quad_epx}
\end{table*}
\begin{table*}
\begin{center}
\begin{tabular}{|l|l|l|l|l|l|l|l|}
\hline 
&&&&&&& \\ 
PROBES & ALGO & LOWER & INIT & HRA & RHRA & REFINED& TIME \\ 
&&BOUND&COST&&&PRECENT& \\ 
\hline \hline 
729 & RAND  &17087 & 26401 & 23970 & 22631 &  14.280\% &  2.83(min) \\ 
729 & SORT  &17087 & 24082 & 22415 & 21649 &  10.103\% &  2.81(min) \\ 
729 & SWM   &17087 & 22267 & 22195 & 22069 &  0.889\% &  2.81(min) \\ 
729 & REPTX &17087 & 21115 & 21107 & 21101 &  0.066\% &  2.81(min) \\ 
729 & EPTX  &17087 & 19733 & 19726 & 19726 &  0.035\% &  2.81(min) \\ 
\hline 
6561 & RAND  &136820 & 243125 & 221090 & 209514 &  13.825\% &  17.55(min) \\ 
6561 & SORT  &136820 & 210326 & 198972 & 191915 &  8.754\% &  17.02(min) \\ 
6561 & SWM   &136820 & 204955 & 204525 & 203412 &  0.753\% &  17.20(min) \\ 
6561 & REPTX &136820 & 185386 & 185362 & 185341 &  0.024\% &  17.16(min) \\ 
6561 & EPTX  &136820 & 168676 & 168623 & 168544 &  0.078\% &  17.15(min) \\ 
\hline 
1024 & RAND  &23480 & 37192 & 35236 & 33046 &  11.148\% &  0.28(sec) \\ 
1024 & SORT  &23480 & 33784 & 32326 & 31026 &  8.164\% &  0.26(sec) \\ 
1024 & SWM   &23480 & 31424 & 31383 & 31323 &  0.321\% &  0.13(sec) \\ 
1024 & QEPX &23480 & 28060 & 28035 & 28028 &  0.114\% &  0.47(sec) \\ 
1024 & REPTX &23480 & 29574 & 29557 & 29546 &  0.095\% &  0.11(sec) \\ 
1024 & EPTX  &23480 & 27591 & 27567 & 27565 &  0.094\% &  0.11(sec) \\ 
\hline 
4096 & RAND  &86818 & 151116 & 143246 & 134485 &  11.005\% &  6.93(sec) \\ 
4096 & SORT  &86818 & 131291 & 127033 & 121742 &  7.273\% &  4.46(sec) \\ 
4096 & SWM   &86818 & 127516 & 127357 & 127092 &  0.333\% &  1.27(sec) \\ 
4096 & QEPX &86818 & 107805 & 107766 & 107702 &  0.096\% &  5.04(sec) \\ 
4096 & REPTX &86818 & 116406 & 116395 & 116376 &  0.026\% &  1.02(sec) \\ 
4096 & EPTX  &86818 & 106471 & 106462 & 106448 &  0.022\% &  1.04(sec) \\ 
\hline 
\end{tabular}

\end{center}
\caption{Cost refinement for various placement heuristics by applying $HRA$ (hierarchical refinement algorithm) and
$RHRA$ (randomized hierarchical refinement algorithm) with $350$ iterations }
\label{table:refinement}
\end{table*}

\subsection{Performance of refinement algorithms}
We have applied our $HRA$, $RHRA$ refining algorithms on the following placement heuristics. 
\begin{itemize}
\item ($RAND$) Random placement: in this placement we just use the order in which the probes
are provided to our algorithm.
\item ($SORT$) Sort placement: in this placement the input probes are sorted lexicographically 
\item ($SWM$) Sliding Window Matching placement is obtained by running the $SWM$~\cite{epitaxial03}
algorithm with parameters $(6,3)$.
\item ($REPX$) Row epitaxial placement is obtained by running the row-epitaxial algorithm with $3$ 
look-ahead rows.
\item ($EPX$) Epitaxial placement is obtained by running the $EPX$ algorithm
\item ($QEPX$) Quad epitaxial placement obtained by our quad-epitaxial algorithm
\end{itemize}
The cost of the placement obtained by running the $HRA$ algorithm exactly once is given in column-5 ($HRA$).
Column-6 ($RHRA$) indicates the placement cost obtained by running our randomized refinement algorithm $RHRA$
for $350$ iterations. From table~\ref{table:refinement} we can see that as initial placement moves closer and
closer towards the lower bound the refinement percentage decreases, which is logical. For test cases with $729$,
$6561$ ($1024$, $4096$) probes we use a refinement degree $d=3$ ($d=2$). Choosing a bigger refinement degree gives
better refinements, however takes more time. Finally we conclude that our refinement algorithms would be very useful
when applied in conjunction with fast initial placement heuristics. A fully function program called {\sf blm-solve}
implementing all our algorithms can be downloaded from the website \url{http://launchpad.net/blm-solve}, the web-site
also has all the supplementary details used in the our experimental study.

\section{Conclusions} \label{sec:conclusion}
In this paper we have studied the Border Length Minimization Problem
(BLMP) that has numerous applications in biology and medicine. We
have solved a seven-year old open problem in this area by showing
that the BLMP is $\cal NP$-hard. Two different proofs have been
given and we believe that the techniques in these proofs will find
independent applications. We have also shown that certain
generalizations of the BLMP are $\cal NP$-hard as well. In addition,
we have presented a hierarchical refinement algorithm (HRA) for the
BLMP. Deterministic and randomized versions of this algorithm can be
used to refine the solutions obtained from any algorithm for solving
the BLMP. Our experimental results indicate that indeed HRA can be
useful in practice.

One of the best performing algorithms for the BLMP is the epitaxial
algorithm (EPX). This algorithm takes too much time especially when
the number of probes is large. In this paper we present a variant
called the quad-epitaxial algorithm (QEPX) that is much faster than
EPX while yielding a solution that is very close to that of EPX in
quality. QEPX partitions the input into four parts, works on each
part separately, and finally combines these solutions. This idea can
be extended further to partition the input into more parts and hence
this algorithm is ideal for parallelism.

Some of the open problems are: 1) In this paper we have used a simple lower
bound on the quality of solution for the BLMP. It will be nice to develop
tighter lower bounds; 2) Develop more efficient algorithms than EPX; and
3) Design parallel algorithms for the BLMP.

\noindent {\bf Acknowledgements.} This work has been supported in
part by the following grants: NSF 0326155, NSF 0829916 and NIH
1R01GM079689-01A1.

\bibliographystyle{abbrv}
\bibliography{vamsi_algos.bib}

\begin{thebibliography}{1}

\bibitem{chatterjee2006}
M.~Chatterjee, S.~Mohapatra, A.~Ionan, G.~Bawa, R.~Ali-Fehmi, X.~Wang,
  J.~Nowak, B.~Ye, F.~A. Nahhas, K.~Lu, S.~S. Witkin, D.~Fishman, A.~Munkarah,
  R.~Morris, N.~K. Levin, N.~N. Shirley, G.~Tromp, J.~Abrams, S.~Draghici, and
  M.~A. Tainsky.
\newblock Diagnostic markers of ovarian cancer by high-throughput antigen
  cloning and detection on arrays.
\newblock {\em Cancer research}, 66(2):1181--1190, 2006.

\bibitem{1.5-approx-HTSP}
N.~Christofides.
\newblock Worst-case analysis of a new heuristic for the travelling salesman
  problem.
\newblock {\em Graduate School of Industrial Administration}, Report 388, 1976.

\bibitem{CAR-RAH-2006a}
S.~de~Carvalho~Jr. and S.~Rahmann.
\newblock Microarray layout as a quadratic assignment problem.
\newblock In {\em Proc. German Conference on Bioinformatics}, volume P-83 of
  {\em Lecture Notes in Informatics}, pages 11--20, 2006.

\bibitem{hannenhalli96}
S.~Hannenhalli, E.~Hubell, R.~Lipshutz, and P.~A. Pevzner.
\newblock Combinatorial algorithms for design of dna arrays.
\newblock {\em Advances in biochemical engineering/biotechnology}, 77:1--19,
  2002.

\bibitem{epitaxial03}
A.~Kahng, I.~Mandoiu, P.~Pevzner, S.~Reda, and A.~Zelikovsky.
\newblock Engineering a scalable placement heuristic for dna probe arrays.
\newblock In {\em Intl. Conf. on Research in Computational Molecular Biology},
  pages 148--156, April 2003.

\bibitem{kundetibibe2009}
V.~Kundeti and S.~Rajasekaran.
\newblock On the hardness of the border length minimization problem.
\newblock In {\em IEEE International conference on bioinformatics and
  bio-engineering}, pages 248--253, 2009.

\bibitem{melle2004}
C.~Melle, G.~Ernst, B.~Schimmel, A.~Bleul, S.~Koscielny, A.~Wiesner,
  R.~Bogumil, U.~Möller, D.~Osterloh, K.~. Halbhuber, and F.~Von~Eggeling.
\newblock A technical triade for proteomic identification and characterization
  of cancer biomarkers.
\newblock {\em Cancer research}, 64(12):4099--4104, 2004.

\bibitem{welsh2003}
J.~B. Welsh, L.~M. Sapinoso, S.~G. Kern, D.~A. Brown, T.~Liu, A.~R. Bauskin,
  R.~L. Ward, N.~J. Hawkins, D.~I. Quinn, P.~J. Russell, R.~L. Sutherland,
  S.~N. Breit, C.~A. Moskaluk, H.~F. Frierson~Jr., and G.~M. Hampton.
\newblock Large-scale delineation of secreted protein biomarkers overexpressed
  in cancer tissue and serum.
\newblock {\em Proceedings of the National Academy of Sciences of the United
  States of America}, 100(6):3410--3415, 2003.

\end{thebibliography}

\end{document}